\documentclass[11pt]{article}

\begin{document}

\title{\Large \bf  Zeeman-like topologies in Special and General theory of Relativity }
\date{}
\maketitle{\noindent\small {{R \ V \ Saraykar $\ ^{1}$ and  Sujatha \ Janardhan$\ ^{2}$}}  \\$\ ^{1}$ {Department of Mathematics,
R\ T\ M Nagpur University, Nagpur-440033. \\ e-mail :  ravindra.saraykar@gmail.com
\\ $\ ^{2}$ {Department of Mathematics, St.Francis De Sales College,  Nagpur-440 006. \\ e-mail :
sujata\_jana@yahoo.com}}}

\vspace{5mm}
\noindent\textbf{Abstract} \ This is a short review article in which we discuss and summarize the works of various researchers over past four decades on Zeeman topology and Zeeman-like topologies, which occur in special and general theory of relativity. We also discuss various properties and inter-relationship of these topologies.

\vspace{5mm}

\section{ Introduction }
\hspace{10mm} \ In special as well as general theory of relativity, space-time models are usually taken as differentiable manifolds. The main reason for representing a space-time as a topological space which is also a differentiable manifold is that we need space-time to have a well-defined topological dimension and we can talk about curves and their tangent vectors, and neighbourhoods to develop a causal theory of space-time. This is achieved by assuming a pseudo-metric structure on a space-time manifold which enables us to define time-like, null and space-like vectors and corresponding curves. In general theory of relativity, metric also determines the geometry and curvature of space-time which represents the gravitational field. In special theory of relativity, Minkowski space M is usually given the topology of real 4-dimensional Euclidean space.

According to Zeeman [1], this topology is not physically reasonable for two reasons: First, the 4-dimensional Euclidean topology is locally homogeneous, whereas Minkowski space M is not,  because to every point in M, there is an associated light cone  which separates space-like vectors from time-like vectors. Secondly, the group of all homeomorphisms of 4-dimensional Euclidean space is vast and is of no physical significance. So, he proposed a new topology for Minkowski space, which is now well-known as Zeeman topology. This is defined as the finest topology on M  which induces 3-dimensional Euclidean topology on every space axis and 1-dimensional Euclidean topology on every time axis. Zeeman proved that this topology has the following physically reasonable properties: Firstly, this topology is not locally homogeneous, and light cone through any point can be derived from the topology. Secondly, the group of all homeomorphisms of this topology is generated by the inhomogeneous  Lorentz group and dilatations.

Zeeman also proved that the topology on a light ray induced from this fine topology is discrete. This means that every function on the light cone is continuous, as every function will be continuous if the domain space has discrete topology. In quantum field theory also, we face similar difficulties regarding 'real' space-time topology, where we talk frequently about continuous wave functions and fields, but we really do not know the meaning of that because 'real' topology of space-time is unknown. However, studies have been dedicated to the topological properties of function spaces, such as spaces of quantum fields, but the study of proper space-time topology which is the most important space of all Physics, remains incomplete. Here, we need a topology in which known quantum quantities such as classical paths on which integrations are to be performed in Feynman's formalism, or Green's functions are continuous. We note that if a function is continuous on a space with topology T, it will be continuous in any refinement of T.

We also note that Zeeman topology is a refinement of $\ E^4 = M $ with 4-dimensional Euclidean topology, but a function which is continuous in Zeeman topology could be discontinuous in the Euclidean topology. Thus, the procedures by which physical quantities such as Green's functions and S-matrix elements, defined on space-time, are transformed by, say, analytic continuation into analogous quantities on  $\ E^4 $, will put constraints on possible topologies on space-time.

On mathematical side, $M$ with Zeeman topology is not a normal topological space , as proved by Dossena [2] and hence it can not be a differentiable manifold, since , by definition, a differentiable manifold is Hausdorff and paracompact as a topological space, and hence normal.

After Zeeman published his paper in 1967, it attracted attention of some of the relativists cum mathematicians and they proved a number of results which are refinements over Zeeman's work. Modified results about Zeeman- and Zeeman-like topologies were published in the context of both special as well as general theory of relativity. Most remarkable are the results by S. Nanda [3,4,5], G. Williams [6], R. G$\ddot{o}$bel [7,8], Hawking-King-McCarty [9], Malament [10] and Lindstrom [11] proved in 1970's. S.G. Popvassilev [12] generalized some of these results to $R^n$. Around 2005, researchers started gaining renewed interest in this field, and further interesting results were published by D.H. Kim [13], G. Dossena [2], G. Agrawal and S. Shrivastava [14,15], G. Agrawal and Soami P. Sinha [16] and R. Low [17]. In fact R. Low extended the results of G. Agrawal and S. Shrivastava [14] to any dimension and also to general curved space-times. He used simpler arguments which do not require the use of Zeno sequences. We reproduce this proof for the sake of completeness.

Since Zeeman topology and other fine topologies defined in special and general theory of relativity in above works have many interesting properties, we discuss these properties and also discuss inter-relationships among these topologies. Most important and remarkable of these results are the results proved by R. G$\ddot{o}$bel and G. Dossena. G$\ddot{o}$bel proved that the group of all homeomorphisms of a space-time of general relativity with Zeeman-like topology is the group of all homothetic transformations. And Dossena proved that the first homotopy group of Zeeman topology for Minkowski space is non-trivial and contains uncountably many subgroups isomorphic to Z. In particular, this topology is not simply connected. Lindstrom generalized the results of G$\ddot{o}$bel and gave a sequence of Zeeman-like topologies which are in the ascending order of fineness.Thus, in Section 2, we describe Zeeman topology and other fine topologies on Minkowski space and discuss their properties. We also discuss t-topology, s-topology and A-topology introduced by Nanda [3-5] and studied in details by G. Agrawal and S. Shrivastava [14,15]. In Section 3, we describe path topology of Hawking-King-McCarty (HKM topology), and improvements by Malament[10], Fullwood [18] and D.H. Kim [13]. We also discuss properties of HKM topology proved recently by R. Low. In Section 4, we describe the work of G$\ddot{o}$bel on Zeeman-like topologies defined on space-time of general relativity and discuss the results proved by him. We also remark on the work of other researchers, especially that by Lindstrom [11] and Mashford [19].

\section{ Zeeman- and Zeeman-like topologies on Minkowski space }
\hspace{10mm} \ We begin this section with definition of \emph{Zeeman topology} as given in Dossena [2]. Let $\ M $ denote 4- dimensional Minkowski space-time and $\ M_{0}$ denote the associated 4- dimensional real vector space equipped with a non-degenerate symmetric bilinear form g of signature ( - , + , + , + ). In $\ M_{0}$, vector axes are either space-like hyperplanes passing through the origin or straight time-like lines passing through the origin. We denote by $\ \mathcal{A}_{0}$ the set of vector axes, and the set $\ p + A_{0}$   with $\ p \in M $ and $\ A_{0}  \in  \mathcal{A}_{0}$, is an axis. We denote the set of axes by $\ \mathcal{A} $. The  Zeeman topology, denoted by $Z$, is by definition, the finest topology on $\ M $ with the property that it induces the affine space natural topology on every axis. $M$ endowed with $Z$ is denoted by $\ M^{Z}$.

A set $U$ is open in $\ M^{Z}$  if and only if for every $\ A \in \mathcal{A},  U \cap A $ is open in $\ A^{E}$. Here, $\ A^{E}$ is the set $\ A $ with natural topology i.e. Euclidean topology. As proved in Zeeman [1] and Dossena [2], the homeomorphism group of $\ M^{Z}$ is generated by the Lorentz group, translations and dilatations.We denote this group by $\ G$.

Physically speaking, the {Zeeman topology $\ M^{Z}$} is defined as the finest topology on a space-time such that its induced topology
on world lines of freely falling test particles with positive rest mass, and on space-like hypersurfaces, is locally Euclidean. Zeeman topology is not as nice as manifold topology, e.g. it is not a normal topological space. On the other hand it has many physically interesting properties: The Zeeman topology does not provide any geometric information along a light ray. Mathematically the topology induced by the Zeeman topology on a light cone is discrete. Secondly, there are many unphysical world lines, e.g. bad trips ( cf Penrose [20] ).

If we interpret continuity of a world line with respect to Zeeman topology, \emph{world lines} are automatically physically realistic, namely, piecewise geodesics which are future directed and time-like with finitely many edges. Hence a world line is the orbit of a freely falling test particle within the gravitational field with a finite number of collisions. This result is a well known basic assumption for a kinetic theory in general relativity ( cf Ehlers [21]).

Moreover if we allow the Zeeman topology to depend on a gravitational field as well as on the Maxwell field, it is possible to derive the corresponding result for charged particles as we discuss below.

In addition to above discussion, we also note that the group of all homeomorphisms of a space-time with its manifold  topology is
neither of interest for physics nor for mathematics since it is vast and it reflects no information of space-time. However, the
group of all homeomorphisms of a space-time $\ M $ with respect to its Zeeman topology $\ M^{Z}$  coincides with its group of all
homothetic transformations, i.e. homeomorphisms are isometries or isometries upto a constant factor. Thus homeomorphisms are proper
symmetry transformations of the space-time. As proved in Zeeman [1], for a Minkowski space, the homothetic transformations are
Lorentz transformations or dilatations of Minkowski space. Hence the homeomorphism group of Minkowski space under Zeeman topology is
its Weyl group, which is generated by Lorentz transformations and linear dilatations.

After Zeeman published his paper in 1967, the first paper by other researcher on this topic was that of S. Nanda [3] in 1971 followed by another one in 1972 [4]. Nanda [3] proved one of the Zeeman's conjecture that the group of homeomorphisms of the finest topology on Minkowski space which induces three dimensional Euclidean topology on every space-like plane is the group $G$. To prove this conjecture, Nanda, like Zeeman, studied chronology preserving and causality preserving mappings and used the notion of Zeno sequences. He defines two topologies, space topology and s-topology with a fine distinction that space topology is strictly finer than s-topology. We recall definitions of these topologies as it would facilitate us to understand other work on fine topologies and compare it with the work of Nanda and Zeeman. As noted above, the space topology on M is defined as the  finest topology with respect to which the induced topology on every space-like hyperplane is Euclidean. Let $M^S$ and $M^E$ denote Minkowski space M equipped with space topology and Euclidean topology. Then space topology is finer than Euclidean topology and hence Hausdorff.\\
Let $\ Q(x) = x_{0}^{2} - x_{1}^{2}- x_{2}^{2}-
x_{3}^{2}$ where $\ x = (x_{0}, x_{1}, x_{2}, x_{3}) \in M $.
Then $Q(x)$ denotes the Minkowski quadratic form. We denote by $\ C^{N}(x)$, $\ C^{T}(x)$ and $\ C^{S}(x)$ the following cones at $\ x$ :\\
Light cone or null cone at $\ x $ : $\ C^{N}(x) = \{ y /   Q (y - x) = 0 \} $ , \\
Time-like cone at $\ x :  C^{T}(x) = \{ y / y = x $ or $\ Q (y - x) > 0 \} $ , \\
Space-like cone at $\ x$ : $\ C^{S}(x) = \{ y / y = x $ or $\ Q (y - x) < 0 \} $ . \\
Let $\ C^{NT}(x) =   C^{N}(x) \cup  C^{T}(x)$. \\
Furthermore, let $\ N^{E}_{\epsilon}(x)$ denote Euclidean $\ \epsilon $-neighbourhood of $\ x $ given by $\ N^{E}_{\epsilon}(x) = \{y /
\rho(x,y) < \epsilon \}, \rho $ being the Euclidean metric, and let \\
$\ N^{s}_{\epsilon}(x) =  N^{E}_{\epsilon}(x) \cap C^{S}(x)$.\\
Then the topology generated by the family $\{ N^{s}_{\epsilon}(x) /{\epsilon} > 0 \}$ of local neighbourhoods at $\ x$ which induces three dimensional Euclidean topology on every space-like hyperplane is s-topology as defined by Nanda, and we denote Minkowski space with this topology by $\ M^s$. Then $\ M^S$ is strictly finer than $\ M^s$. After proving a series of lemmas about chronology preserving homeomorphisms, Nanda [3] proves that the group of homeomorphisms of $\ M^S$ is $G$.
In the subsequent paper, Nanda [4] defines t-topology in a similar way :\\
Let $\ N^{t}_{\epsilon}(x)$ = $\ N^{E}_{\epsilon}(x) \cap C^{T}(x)$.\\
Then t-topology is defined as the topology which has the family $\{ N^{t}_{\epsilon}(x) /{\epsilon} > 0 \}$ as a local base of neighbourhoods at each point $\ x$ of $\ M$. $\ M$ equipped with this topology is denoted by $\ M^t$. In [4], Nanda proves another version of Zeeman's conjecture, namely that the group of homeomorphisms of $\ M^t$ is the group $\ G$. ( Theorem 1 [4]). Furthermore, he also proves that the group of homeomorphisms of $\ M^s$ is also group $\ G$ (Theorem 2 [4]).
If $\ M$ and $\ M'$ are Minkowski spaces, ( or space-times of general relativity ) then the mapping $\ f : M \rightarrow M' $
with the property that both $\ f$ and $\ f^{-1}$ preserve chronological order is known in the literature as chronal isomorphism (cf. P.S. Joshi [22]) Similarly, if both $\ f$ and $\ f^{-1}$ preserve causal order, then f is called causal isomorphism or simply a causal map. Such maps are extensively studied in the literature as cone preserving mappings ( see for example, Garcia-Parrado and Senovilla [23] and S. Janardhan and R.V. Saraykar[24], and references therein ).

Williams [6] studies other Zeeman-like topologies on the Minkowski space and  derives homeomorphism groups for these topologies. We summarize below the results proved by Williams. It is interesting to note that $\ C^1$ subgroup of homeomorphism group of some of these fine topologies is the same as $\ G$.

Here $\ M^{E}$ is $\ M $ with natural topology as above and so are $\ C^{N}(x)$, $\ C^{T}(x)$ and $\ C^{S}(x)$. Let $\ {M^{F_{i}}} $ denote the set of finest topologies such that the restrictions of the identity mapping of $\ M^{E} $ onto each $\ {M^{F_{i}}} $ to time-like and space-like lines are homeomorphisms. Williams proves that there is a unique such finest topology. $\ M $ with this topology is denoted by $\ M^{F}$. The fine topology $\ M^{F}$ is defined as follows :\\
\noindent{Topology $\ M^F $} is the topology on $\ M $ generated by the local base of open neighbourhoods $\{ N^{F}_{\epsilon}(x), x \in M , \epsilon > 0 \}$ at $\ x$. Here  $\ N^{F}_{\epsilon}(x) $ is defined as \\
$\ N^{F}_{\epsilon}(x)\  =  N^{E}_{\epsilon}(x)\cap  C^{ST}(x)$ where  $\ C^{ST}(x) =   C^{S}(x) \cup  C^{T}(x)$. \\
He further proves that the group of homeomorphisms of $\ M^{F}$ is the conformal group of Minkowski space. This is in fact the group generated by the Lorentz group, translations and dilatations, and thus, it is the same as $\ G$.\\
Williams further describes two more fine topologies for $\ M $ and describes their homeomorphism groups. The first of these topologies is $\ M^{T}$. A physically significant topology for $\ M $ is the finest topology such that the restrictions of the identity mapping of $\ M^{E} $ onto $\ {M^{F_{i}}} $ to time-like lines are homeomorphisms. In this topology the relative topology along space-like lines is discrete. $\ M^{T} $ is Minkowski space with this fine topology. Group of $\ C^{1}$ - homeomorphisms of $\ M^{T}$ is the conformal group which is again same as $\ G$.

Following the argument in Nanda [3], though it can be proved that $\ M^{T} $ is strictly finer than t-topology, homeomorphism groups of both these topologies are the same and their topological properties are also similar.

Second of these topologies is $\ M^{L}$.  $\ M^{L} $ is the unique finest topology such that the restrictions of the identity mapping of $\ M^{E} $ onto $\ M^{L} $ to straight lines are homeomorphisms. Also, there exists a unique such finest topology and that it is strictly finer than $\ M^{E} $. It is weaker than the two previous topologies discussed here. The line sequence introduced here is however a Zeno sequence and any homeomorphic image of $\ I $ must be piecewise linear.   (Here $\ I $ is the closed unit interval). Thus the group of $\ C^{1} $ homeomorphisms of $\ M^{L} $ does preserve straight lines and
is thus a subgroup of the projective group on $\ R^{4} $. In fact, group of $\ C^{1} $ - homeomorphisms of $\ M^{L}$  is the projective group which is generated by full linear group and translations. Thus it coincides with homeomorphism group of $\ M^{E}$. This work resembles the work of S. Nanda [3,4,5]. In fact, in the third paper [5], Nanda defines yet another fine topology on the Minkowski space, called A-topology and derives its homeomorphism group. He also compares his results with those of Williams. The A-topology is defined as follows : \\
\textbf{Definition 2.1 : A-topology}:  The A-topology on $\ M $ is defined to be the finest topology on $\ M $ with respect to which the induced topology on every time-like line and light-like line is one-dimensional Euclidean and the induced topology on every space-like hyperplane is three-dimensional Euclidean.\\Thus A-topology is strictly finer than the Euclidean topology.

The topology $\ M^{F}$ suggested by Williams on Minkowski space is characterized by the property that the induced topology on time-like and space-like lines is Euclidean and that it is the finest such topology on $\ M $ having this property. This topology differs significantly from the A-topology (or from Zeeman's fine topology) in its group of homeomorphisms. Williams has proved that the $\ C^{1}$-subgroup of homeomorphisms of this topology is $\ G$. Without the $\ C^{1}$-condition, the result may not be valid. Nanda proves, by using Zeno sequence method, that the group of homeomorphisms of A-topology is also same as $\ G$. Furthermore, as remarked by Nanda [5], if $\ f : I \rightarrow M^F $ is a continuous map, then $\ f(I)$ is a connected union of time-like and (or) space-like intervals. This is in contrast with the result for A-topology where $\ f(I)$ is a connected union of finite number of time-like and (or) null intervals. If, however, $\ f$ is assumed to be order-preserving, then it follows that $\ f(I)$ is a connected union of time-like intervals representing the path of an inertial particle under a finite number of collisions. This excludes the path of photons. Thus A-topology is significantly different from William's topology in this respect.

Popvassilev [12] generalized the concept of Zeeman - like fine  topologies to $\ R^{n}$ and proved that these topologies are non-regular. Since these topologies are Hausdorff, it follows that they are not normal. This property was proved by Dossena [2] in a different way by using Urysohn Lemma.

S. Nanda and H.K. Panda [25] define yet another topology on Minkowski space. This is a non-Euclidean topology, namely order topology generated by the positive cone at origin and its translates. They prove that it is non-compact, non-Hausdorff but path-wise connected.  Moreover, it has the property that every loop based at a point is homotopic to the constant loop at that point. Thus, this topology is simply-connected. This is contrary to the non-simply connected nature of $\ M^Z$, $\ M^t$ and $\ M^s$.

 We now discuss the work of Dossena[2] and G. Agrawal and S. Shrivastava[14,15] where many interesting topological properties of $\ M^Z$, $\ M^t$ and $\ M^s$ have been proved, including non-simply connectedness when restricted to two dimensional Minkowski space.

As defined in the begining of this section, Dossena presents Zeeman topology $\ M^{Z}$  in the language of affine spaces and proves that Zeeman topology is separable, non- first countable and non-trivial. We discuss below the results proved by Dossena in some details, especially for two dimensional Minkowski space.

For two dimensional Minkowski space with topologies $\ M^{E}$ and $\ M^{Z}$, Dossena gives characterization of the sets $\ \Sigma \subseteq M$ on which $\ M^{E}$ and $\ M^{Z}$ induce the same topology. i.e. $ \ \Sigma \cap M^{E} =  \Sigma \cap M^{Z}$. To prove this, he uses the concept of Zeno sequences. Furthermore, in this two dimensional case, he gives characterization of compact subsets of $\ M^{Z}$. We summarize these
results below :\\
\textbf{Lemma 2.1}\ A compact subset of $\ M^{Z}$ is compact in $\ M^{E}$.\\
\textbf{Lemma 2.2}\ Let X be a Hausdorff topological space and let $\ (x_{n})_{n \in N}$ be a sequence of
distinct points of X converging to x. Then x is the unique limit point for the set $\ \{x_{n}\}_{n \in N}$ .
In particular, every $\ x_{j} $ is an isolated point for $\ \{x_{n}\}_{n \in N}$.\\
\textbf{Lemma 2.3}\ Every Zeno sequence admits a subsequence whose image is a non closed,
discrete subset of $\ M^{E}$, closed in $\ M^{Z}$.\\
\textbf{Theorem 2.4}\ A compact subset K of $\ M^{Z}$ contains no images of Zeno sequences.\\
This is true for A-topology also, as proved by Nanda [5]. \\
\textbf{Theorem 2.5 } For a subset $\ K \subset M $, the following are equivalent:\\
1. K is compact in $\ M^{Z}$.\\
2. K is compact in $\ M^{E}$ and contains no completed images of Zeno sequences.\\
3. K is covered by a finite family $\ (A_{j})_{j=1,...,J} $ of axes such that for each j = 1, . . . , J the
set $\ A_{j}\cap K $ is compact in $\ A^{E}_{j} $.\\
We now discuss countability properties of $\ M^{Z}$. \\
We choose  an orthonormal frame of reference $\ (o, (e_{i})_{i=0,...,k} $). Then every $\ p \in M $ is
identified by its coordinates $\ \{p^{i}\}_{i=0,...,k} $, such that $\ p = o + \sum^{k}_{i=0}  p^{i}e_{i}$. \\
 Clearly $\ M^{E}$ is separable (so are all finite-dimensional affine spaces endowed with their
natural topology). A countable dense subset Q of $\ M^{E}$ can be constructed by choosing
an orthonormal frame of reference and defining Q as the set of points in M with rational
coordinates.\\
Then we have the following proposition:\\
\textbf{Proposition 2.6} \  For every orthonormal frame of reference, the above-mentioned set Q is also dense in $\ M^{Z}$. Thus $\ M^{Z}$ is separable.\\
\textbf{Corollary 2.7} \ The cardinality of the set $\ \mathcal{C}(M^{Z},R)$ of all real continuous functions on
$\ M^{Z}$ is at most equal to $\ 2^{\chi_{0}}$, where $\ \chi_{0}$ is the cardinality of Natural numbers.\\
\textbf{Proposition 2.8}\  $\ M^{Z}$ is not first countable at any point.\\
Zeeman [1] has sketched the proof of the result that  $\ M^{Z}$ is not normal. As noted earlier, Dossena gives another proof of the same result using Urysohn lemma. Thus, we have :\\
\textbf{Theorem 2.9}\  $\ M^{Z}$ is not normal and hence  not metrizable.\\
For a path-connected topological space $X$, $\ \pi_{1}({X})$ denotes the fundamental group or first homotopy group of $X$.
The following is the most remarkable result proved by Dossena :\\
\textbf{Theorem 2.10}: $\ \pi_{1}(M^{Z})$ is nontrivial and possesses uncountably many subgroups isomorphic to Z. In particular, $\ M^{Z}$ is not simply connected.
For details of proofs we refer the reader to Dosssena [2].

A topological study of the n-dimensional Minkowski space, $\ M^ {n}$, with t-topology, denoted by $\ M^t$, has been carried out by G. Agrawal and S. Shrivastava[14]. Path-topology defined by Hawking, King and Mc Carthy [9] on a space-time of general relativity will be discussed in Section 3. If we restrict this topology to four dimensional Minkowski space, then it comes out to be identical with t-topology. Non-simply connectedness of $\ M^{t}$, compact sets of $\ M^{t}$, and subsets of $M$ that have the same subspace topologies induced from the Euclidean and t-topologies are also discussed in this paper.\\
t-topology for four dimensional Minkowski space has been defined above. Similar definition follows for $\ M^ {n}$ also. Thus $\ U \subset M$ is open with respect to t-topology if and only if for each $\ x \in U $ there exists some $\ N^{t}_{\epsilon}(x)$ such that $\ N^{t}_{\epsilon}(x) \subset U $.\\
It  thus follows that $\ N^{E}_{\epsilon}(x)$  and $\ N^{t}_{\epsilon}(x)$ are open in M with t-topology, $\ N^{E}_{\epsilon}(x)$ is open in M with Euclidean topology, while $\ N^{t}_{\epsilon}(x)$  is not open in $M$ with Euclidean topology. Hence $\{ N^{t}_{\epsilon}(x) /\ x \in M,{\epsilon} > 0 \}$ is a basis for the t-topology and the t-topology is strictly finer than the Euclidean topology on M. \\ s-topology can be defined similarly on $\ R^{n}$. \\
Summarizing, we have the following : \\
The collection $\{ N^{t}_{\epsilon}(x) /\ x \in M, {\epsilon} > 0 \}$  being a basis for the path topology on four-dimensional Minkowski space, the path topology on four-dimensional Minkowski space is same as the t-topology. It thus follows that the four-dimensional Minkowski space with t-topology is Hausdorff, path connected, separable, first countable, not second countable, not countably compact, not Lindelof, not regular, not normal and hence is not compact, not locally compact, not paracompact, not mertizable, and not locally n-Euclidean.

Other works on Zeeman-like topologies include that of Struchiner and  Rosa [26] and Domiaty [27,28]:\\
Struchiner and  Rosa [26] study Zeeman topology in Kaluza-Klein and Gauge theories. They generalize the notion of Zeeman topology by using the projection theorem of Kaluza -Klein theories, and this remains valid for any gauge fields. Here, the authors consider differential geometric frame work of fiber bundles and define Zeeman topology in the total space of fiber bundle. From this, they obtain a topology in the base manifold for which the continuous curves correspond to motions of charged particles in the base manifold. It would be interesting to see the generalizations of typical gauge theoretical ideas when the space-time has such a topology.\\
Domiaty [27,28] considers yet another topology on Lorentz manifolds. This topology is in a certain sense the space-like version of an analogous result for the Hawking-King-McCarthy path topology which has been discussed below. The space topology is the finest topology on a Lorentz manifold, which induces the manifold topology on every space-like hypersurface. As proved in these papers, its geometric significance comes from the fact that its full homeomorphism group is the group of all conformal diffeomorphisms. \\
Finally, we remark that even though Zeeman topology on Minkowski space has several advantages over the standard topology, it has some drawbacks  also. These are as follows: \\
(i)  A three dimensional section of simultaneity has no meaning in terms of physically possible experiments. Also, the use of
straight time like lines in defining $\ M^{Z}$ suggests that $\ M^{Z}$ from the beginning has been equipped with information
involving inertial observers, so that occurrence of linear structure is not surprising. \\ (ii)  The isometry and conformal groups of $\ M^{Z}$ are physically significant but same thing is not clear about homothecy group of $\ M^{Z}$ \\
(iii)   The set of $\ M^{Z}$- continuous paths does not incorporate accelerating particles moving under forces in curved
lines. \\ iv)  $\ M^{Z}$ is not first countable and hence it is difficult to handle.\\
Keeping these drawbacks in mind, Hawking, King and Mc Carthy [9] defined another topology called path topology on a space-time of
general relativity. We now discuss, below, this topology and its properties. We also discuss other related topologies as studied by Kim [13] and Low [17] and their inter-relationships with HKM topology.

\section{Path topology of Hawking, King and Mc Carthy (HKM) and other related topologies}
 Here, we consider a space-time of general relativity which is assumed to be connected, Hausdorff, paracompact, $\ C^{\infty} $ real four- dimensional manifold $\ V $ without boundary, with a $\ C^{\infty} $ - Lorentz metric and associated pseudo - Riemannian connection. $\ V $ is also assumed to be time-orientable i.e. $\ V $ admits a non-vanishing time-like vector field. \\
The path topology $\ \mathcal{P}  $ of $\ V $ is defined as follows:\\
$\ \mathcal{P}  $ is the finest topology satisfying the requirement that the induced topology on every time-like curve coincides with the topology induced from $\ \mathcal{V}  $, where $\ \mathcal{V}  $ is the given manifold topology on $\ V $.

Thus if a set $\ E \subset  V $ is $\ \mathcal{P}  $ - open, for every time-like curve $\ \gamma $, there is an $\ O \in
\mathcal{V} $ with \  $\ E \cap \gamma = O \cap \gamma $. Conversely, if E satisfies this condition, it is $\ \mathcal{P} $- open and $\ \mathcal{P} $ is the largest collection of such sets. Obviously, if $\ O \in   \mathcal{V} $  , then $\ O \in \mathcal{P} $.\\
HKM show that $ \mathcal{P} $ is strictly finer than $ \mathcal{V} $, but however $ \mathcal{P} $ is not comparable to Zeeman
topology.

Let $\ T_p(V)$ denote the tangent space of $\ p \in {V}$ and $\ exp : T_p(V)  \rightarrow V $ be the exponential mapping. Then there is an open neighbourhood $N$ of the origin of $\ T_p(V)$ such that $U = exp (N)$ is an open convex neighbourhood of $\ p \in {V}$. Let $\ \epsilon > 0$ be sufficiently small so that the Euclidean open ball $B$ of radius $\ \epsilon, $ with centre at origin, is contained in $N$. Then $B_u{(p, \epsilon)} = exp (B)$. For any open set $\ V $, define $\ C(p,V) = I^{+}(p,V) \cup I^{-}(p,V), $ \\
$\ K(p,V) = C(p,V) \cup \{p\}$ and for an open convex normal neighbourhood \emph{U} of $\ p $, define \\
$\ L_{u}(p,\epsilon) = B_{u}(p,\epsilon) \cap K(p,U)$. ( $\ B_{u}(p, \epsilon) = exp B $ ). Then we have the following :

\noindent\textbf{Proposition 3.1:} Sets of the form $\ K(p),
K(p,U)$ and $\ L_{u}(p, \epsilon) $ are $ \mathcal{P} $-open.\\
$\ K(p,U) $ is not open in the manifold topology $ \mathcal{V} $ because $\ p \in K(p,U)$ has no $ \mathcal{V} $-nbd contained in
$\ K(p,U) $. Thus $ \mathcal{P} $ is strictly finer than $ \mathcal{V} $.

\noindent\textbf{Theorem 3.2:} $\ L_{u}(p,\epsilon)$ forms a
basis for the topology $ \mathcal{P} $.\\
This property has no analogue in the finer topologies $\ \mathcal{T}$. $ \mathcal{P} $-continuous paths are characterized
as follows:

\noindent\textbf{Theorem 3.3:} A path $\ \gamma : F \rightarrow V
$ is $ \mathcal{P} $ -continuous if and only if it is a Feynman Path.

\noindent\textbf{Theorem 3.4:} $ \mathcal{P} $ is first countable and separable. $ \mathcal{P} $ is Hausdorff, path connected and locally path
connected and hence locally connected. However, $ \mathcal{P} $ is not regular, normal, locally compact or paracompact.\\
Furthermore, HKM determine the group of $ \mathcal{P} $-homeomorphisms and prove that it is the group of smooth conformal diffeomorphisms.\\
To begin with, they prove the following :

\noindent\textbf{ Proposition 3.5:} $ \mathcal{P} $-homeomorphisms h take time-like curves to time-like curves.\\
This has been proved for strongly causal space-times. It is done by
singling out a subclass of $ \mathcal{P} $-continuous curves which
coincides with time-like curves. \\
After proving a series of results, HKM prove the following important theorem :

\noindent\textbf{Theorem 3.6:} A $ \mathcal{P} $-homeomorphism h is a smooth conformal diffeomorphism.
This leads to the description of $ \mathcal{P}$-homeomorphisms of $M$ :

\noindent\textbf{Theorem 3.7:} The group of $ \mathcal{P} $-homeomorphisms of $M$ coincides with the group of  smooth conformal diffeomorphisms of $M$.

Finally, HKM give an example of a manifold for which the group of  smooth conformal diffeomorphisms
is strictly larger than the homothecy group.  We note here that for Minkowski space, the two groups are equal.\\
For more details and proofs, we refer the reader to HKM [9].\\
Malament [10] improved the results of [9] in the sense that the condition of strong causality is no longer
 necessary. We now discuss briefly the work of Malament [10] : \\
Main result of this paper is the following:\\
Suppose we consider two space-times $\ (V,g)$ and $\ (V^{'},g^{'})
$ and a bijection $\ f : V \rightarrow \ V^{'}$, where both $\ f $
and $\ f^{-1}$ preserve continuous time-like curves. This means, if
$\ \gamma : I \rightarrow V $ is a continuous time-like curve in $\
(V,g) $  then $\ f \circ \gamma : I \rightarrow V^{'} $ is a
continuous time-like curve in $\ (V^{'},g^{'}) $. Similar condition
holds for $\ f^{-1}$. Then $\ f $ must be homeomorphism. Thus the
class of continuous time-like curves in a space-time determines its
topology.By Hawking's theorem, $\ f $ will then be a smooth conformal
isometry.\\
Brief summary of the proof is as follows:\\
If $\ f $ preserves all continuous curves, then $\ f $ would be
continuous. Given any sequence $\ \{p_{n} \}$ converging to $\ p
$, one could find a continuous curve 'threading' all the $\ p_{n}
$ in sequence and then $\ p $. Its image would have to be a
continuous curve threading all the $\ f(p_{n}) $ in sequence and
then $\ f(p)$. Hence $\ f( \{p_{n}\}) $ would converge to $\ f(p)
$. Under the hypotheses under consideration, this construction can
only be applied to sequence $\ \{p_{n} \}$ which converge
chronologically to $\ p $. The problem is with those sequences $\
\{p_{n} \}$ which converge to $\ p $ but are locally space-like
related to $\ p $.\\
The idea to overcome this difficulty is as follows:\\
To show that $\ f $ is continuous at $\ p $, one proves that one
may assume that $\ f $ is continuous over a `nice-looking' region
near $\ p $. Then one uses continuous null geodesic segments in
this region to characterize the convergence of points to $\ p $.
This then leads one to the required result because continuous null
geodesics in this region are necessarily preserved by $\ f $. For
technical details, we refer the reader to Malament [10]. HKM - topology is an improvement over Zeeman topologies in the sense that it removes many unpleasant features of those topologies.\\
Fullwood [18] modified the HKM topology and defined a new topology $\ \overline\mathcal{P}$ as follows :\\
$q \gg p$  if and only if $q \in I^+{(p)} $. Then $ p \in I^- {(q)}$ and $p \ll q$.
We denote $\ I^+{(q)} \cup  I^-{(q)}$ by  $\ I(q)$.\\
Then, define $\ <p, q, r> = I^+ {(p)} \cap I(q) \cap I^- {(r)} \cup \{q\}$ for $\ p \ll q \ll r$ in $V $.\\
Now, let  $\ \mathcal{B}  = \{ B : B = <p, q, r> $ for some $\ p \ll q \ll r$ in $V \}$.\\
Then,  $\ \mathcal{B}$  forms  a  base for  a  topology which is denoted by $\ \overline \mathcal{P}$.

Fullwood proves that if the space-time $V$ is future and past distinguishing, then the topology $\ \overline\mathcal{P}$ coincides with HKM $ \mathcal{P}$-topology. More precisely, he proves the following theorem :\\
\noindent\textbf{Theorem 3.8:} The following three conditions are equivalent upon a space-time manifold: \\
(a)$\ \overline\mathcal{P} =  \mathcal{P} $ i.e., the topology $\ \overline\mathcal{P}$ is equivalent to the Path topology;
(b) the distinguishing condition holds on $V$, and (c) $V$ is $\ \overline\mathcal{P}$- Hausdorff.

Do-Hyung Kim [13] proved that the path topology of Hawking, King, and McCarthy can be extended to the causal completion of a globally hyperbolic Lorentzian manifold. The suggested topology $\mathcal{T}$ is defined only in terms of chronological structures and $\mathcal{T}$ is finer than the extended Alexandrov topology denoted by $\ \overline\mathcal{A}$. It is also shown that a $\mathcal{T}$-homeomorphism induces a conformal isomorphism and a homeomorphism in $\ \overline\mathcal{A}$. Let $\ \overline{V }$ denote causal completion of $V$. Then $\mathcal{T}$ is defined on $\ \overline{V }$ as follows:\\
\textbf{Definition 3.1}: $\ U \subset \overline{V } $ is $\mathcal{T}$-closed if every time-like sequence that converges has a limit in $U$ and $\ W \subset \overline{V } $ is $\mathcal{T}$-open if its complement is $\mathcal{T}$-closed.\\
\textbf{Proposition 3.9}: The above family of open sets define a new topology $\ \mathcal{T}$ on $\ \overline{V }$.\\
\textbf{Proposition 3.10}: The topology $\mathcal{T}$ on $\overline{V}$  is finer than the extended Alexandrov topology $\ \overline\mathcal{A}$ on $\ \overline{V}$.\\
Since $\mathcal{T}$ is finer than $\ \overline\mathcal{P}$ and $\ \overline\mathcal{P}$ is Hausdorff, it can be concluded that $\ \mathcal{T}$ is a Hausdorff topology on $\ \overline{V}$.\\
\textbf{Corollary  3.11}: $\ I^{-}(\gamma) $ is also an end point of a time-like curve  $\ \gamma : (a,b) \rightarrow V $  in $\ \mathcal{T}$-topology.\\
The construction of $\mathcal{T}$-topology on the causal completion extends the $\mathcal{P}$-topology on $V$ by use of the sequential convergence.\\
Furthermore, Kim studies homeomorphisms with respect to topology $\mathcal{T}$. To understand the results in this direction, let $V$ and $N$ be two space-times and let $\ \overline{V }$ and $\ \overline{N }$ be their causal completions. Then we have the following definition : \\
\textbf{Definition  3.2}: A bijection $\ f : \overline{V}\rightarrow \overline{N}$  is a chronological isomorphism if $\ x \ll y \Leftrightarrow f(x) \ll f(y) $ and antichronological isomorphism if $\ x \ll y \Leftrightarrow f(y) \ll f(x) $. Likewise, a bijection $\ f : \overline{V } \rightarrow \overline{N}$ is a causal isomorphism if $\ x \leq y \Leftrightarrow f(x) \leq f(y) $ and anticausal isomorphism if $\ x \leq y \Leftrightarrow f(y) \leq f(x) $. A bijection $\ f : \overline{V}\rightarrow \overline{N}$ is a conformal isomorphism if $f$ is both (anti) chronological isomorphism and (anti) causal isomorphism.
In a Lorentzian manifold, it is known that the causal isomorphism and the chronological isomorphism are equivalent.
The topology $\ \mathcal{T}$ is defined only in terms of chronological relations and so any chronological isomorphism $\ f : \overline{V} \rightarrow \overline{N}$ induces a $\ \mathcal{T}$-homeomorphism. The chronological isomorphism has the same effects on the $\ \overline \mathcal{A}$ -topology.We also have the following :\\
\textbf{Proposition 3.12} : If $V$ and $N$ are globally hyperbolic and $\ f : \overline{V}\rightarrow \overline{N}$ is either a chronological isomorphism or an antichronological isomorphism, then $f$ is an $\ \overline\mathcal{A}$ -homeomorphism.\\
\textbf{Theorem  3.13} : If $\ f : \overline{V} \rightarrow \overline{N}$ is a $\ \mathcal{T}$-homeomorphism, then $f$ is either a chronological isomorphism or an antichronological isomorphism.\\
\textbf{Theorem 3.14}: If $\ f : \overline{V} \rightarrow \overline{N}$ is a $\ \mathcal{T}$-homeomorphism, then $f$ is a conformal isomorphism.\\
Since $\ \mathcal{T}$ is finer than $\ \overline \mathcal{A}$ , by combining proposition 3.8 and theorem 3.9, we have the following theorem.\\
\textbf{Theorem  3.15} \ A $\ \mathcal{T}$-homeomorphism induces an $\ \overline \mathcal{A}$ -homeomorphism.\\
Also if $\ f : V \rightarrow N$ is a $\ \overline \mathcal{P}$ -homeomorphism, then $f$ is a conformal isomorphism. If, in
addition, both $V$ and $N$ are strongly causal, the manifold topologies are the same as the Alexandrov topologies since the Alexandrov topology is defined only in terms of a chronological relation. In other words, a $\ \overline\mathcal{P}$-homeomorphism induces an $\ \mathcal{V}$-homeomorphism. By the above theorem, this is indeed the case in the path topology of the causal completion. Thus,  the extended Alexandrov topology is natural to the causal completion. The $\ \overline \mathcal{P}$- topology mentioned here is that defined in Fullwood [18], and the causal completion of space-times mentioned in the discussion above is in the sense of Budic and Sachs [29].

Such bijective mappings have also been studied by Domiaty [27,28]. These mappings are defined in such a manner that they leave the class of space-like paths invariant. Homeomorphisms with respect to $S$ - topology defined by Nanda [4] are called $S$ - homeomorphisms. Domiaty proved that if $(V,g)$ and $(V^{'},g^{'})$ are Lorentz manifolds and if $f : V \rightarrow V^{'}$ is a bijection, then $f$ is a $S$- homeomorphism if and only if $f$ and $\ f^{-1}$ preserve space-like paths. Furthermore, after proving a series of lemmas, he proves that if $f$ and $\ f^{-1}$ preserve space-like paths,then $f$ is a manifold-homeomorphism ( $V$-homeomorphism ). There is a substantial literature on causality-preserving maps (causal maps) or cone-preserving maps in special as well as general theory of relativity. See, for example, a review article by Sujatha Janardhan and R.V. Saraykar [24] and references therein. If we denote homeomorphisms with respect to path-topology (HKM - topology ) by $\mathcal{T}$-homeomorphisms, then every $S$-homeomorphism is a $\mathcal{T}$-homeomorphism. Since (ref. Kim [13]) a $\mathcal{T}$-homeomorphism is a smooth conformal diffeomorphism, it follows, by combining results of Domiaty and Kim, that every $S$-homeomorphism is also a smooth conformal diffeomorphism. (This has been noted by Domiaty [27,28] Theorem 2 ). This result improves the result by G$\ddot{o}$bel [7] which was proved for strongly causal Lorentz manifolds.

More recently Huang [30] proved the result: Let $\ (V , g) $ be a strongly causal space-time, dim $\ V \geq 3$. Let $\ f: V \rightarrow V $ be a bijection such that images and pre-images of null geodesics ( as point sets) are null geodesics. Then f is a homeomorphism and hence by Hawking's theorem,  a conformal transformation. This generalizes the result proved by Jan Peleska [31]. Define a local distance function on convex normal neighbourhoods by $\ \phi ( p , q ) = g ( exp_{p}^{-1} q , exp_{p}^{-1} q )$ Then every homeomorphism f which locally preserves these functions is an isometry. If $\ ( V , g )$ has indefinite signature and f locally preserves distance zero, then it is a conformal diffeomorphism.

The physical meaning of the condition used in this theorem is that images and pre-images of paths which photons travel between emission and absorption should again be such paths.

Coming to the topological properties of Zeeman-like topologies on Minkowski space $M$ again, we note the Theorem proved by Dossena, namely, two dimensional Minkowski space is not simply connected. Its first homotopy group contains uncountably many subgroups isomorphic to $\ Z $. G. Agrawal and S. Shrivastava [13] proved similar result for t-topology. Both these proofs use the notion of Zeno sequences introduced by Zeeman. Robert Low [17] recently gave a proof for the same result for n-dimensional Minkowski space with Zeeman topology without using Zeno sequences. For the sake of completeness, we reproduce the proof of this important theorem below. \\
\noindent\textbf{Theorem  3.16:} \ A space-time $V$, equipped with the path topology is not simply connected or locally simply connected. Furthermore, no two closed continuous curves in $V$ with distinct images are homotopic.\\
\noindent\textbf{Proof}:\ Let $\ c_{1}$ and $\ c_{2}$ be curves in $V$ with distinct images, let $\ h : I \times I \rightarrow V$
such that $\ h(0, .)$ is $\ c_{1}$ and $\ h(1, .)$ is $\ c_{2}$ , and let $T$ be some time-like two-plane
and  $\ \pi $ be the associated projection such that the projections of $\ c_{1}$ and $\ c_{2}$  to $T$ are
distinct. Now neither of $\ \pi o c_{1}$ nor $\ \pi o c_{2}$ can be space-filling, for then we already have an open set in $T$ containing infinitely many points in some space-like surface and in the image of $\ \pi o h $. R. Low then considers the intersection of this open set with some surface of constant time and argues to conclude that there must be some point $x$ in $T$ round which $\ c_{1}$ and $\ c_{2}$ have different winding numbers. Since $\ c_{1}$ and $\ c_{2}$ are closed curves in $T$, $x$ has
an open neighbourhood in $T$ which lies in the image of $\ \pi o h $, and again we obtain a contradiction. Hence, if $\ c_{1}$ and $\ c_{2}$ are closed continuous maps from $\ S^{1}$ to $V$ with distinct images, then $\ c_{1}$ and $\ c_{2}$ are not homotopic in the path topology. Moreover the fundamental group of $V$ with the path topology is as large as possible, since two continuous loops are only homotopic if one is
a re-parameterisation of the other. Also, the above result is true in case of a general Lorentz manifold. The general space-time $V$ can be embeded in a pseudo-Euclidean space of appropriate dimension, and arguing as above, by projecting to some suitable time-like plane in the pseudo-Euclidean space, we can obtain the same result.

Here, it will not be out of place to mention that Sorkin and Woolgar [32] introduced the concept of K-causality with the aim that it should be possible to derive the causal structure from order relation and topological structure. Some results in this direction were proved by  S. Janardhan and R.V. Saraykar [33]. Later, after a good deal of efforts, Minguzzi [34] proved that \emph{Stable causality} is equivalent to \emph{K-causality}. In the description of path-topology above, if analogously, if we replace a time-like curve by a K -causal curve which is compact,connected and linearly ordered, then we can define \emph{K-causal topology} on $\ V $ , denoted by $\ \mathcal{K} $ as follows:\\
We specify closed sets of $\ \mathcal{K} $ as follows:\\
$\ \widetilde{F}$ is a $\ \mathcal{K} $ - closed subset of $\ V $ if \ $\ \widetilde{F} \cap \gamma = F \cap \gamma $ for some
closed $\ F \subseteq V $, in the manifold topology and $\ \mathcal{K} $ is the finest such topology. If F is closed in $\ V $, with respect to $ \mathcal{V} $, then F is closed with respect to $\ \mathcal{K} $ also). Thus $ \mathcal{K} $ is finer than $\ \mathcal{V} $.  For a detailed discussion of $K$-causal curves in $K$ - causal space-time, we refer the reader to S. Janardhan and R.V. Saraykar [31] and Minguzzi [32] and references therein.

\section{Zeeman-like topologies in General Relativity}
In this section, we describe and discuss the work of G$\ddot{o}$bel [7,8], Lindstrom [11] and others on Zeeman-like topologies defined on a space-time of general relativity. In particular, G$\ddot{o}$bel [7] has proved the result that two space-times  are homeomorphic with
respect to its Zeeman topology if and only if they are isometric. This shows that it is possible to determine the metric of a space-time from
its Zeeman topology.

We start with definitions of Zeeman topologies as given by G$\ddot{o}$bel[7] and discuss their main properties.\\
Let  $\ (V, \mathcal{T} $) denote a differentiable manifold with
an underlying manifold topology $\ \mathcal{T} $. The most general
setting for Zeeman topologies is the following: \\
Let $\ \sum $ be a set of subsets of $\ V $.   Then a subset $\ X
\subset V $ belongs to $\ Z = Z (\Sigma, \mathcal{T}) $ \ iff \ $\
X \cap Y $ is open within the topological space $\ Y = ( Y,
\mathcal{T}_{Y}) $ with its induced topology $\ \mathcal{T}_{Y}$,
for all $\ Y \in \Sigma $. -------($\star $) \\
Then $\ (V, Z) $ is the space $\ V $ provided with the Zeeman
topology $\ Z = Z (\Sigma, \mathcal{T}) $ generated by $\ (\Sigma,
\mathcal{T}) $. Thus the topology Z is the finest topology $\
\mathcal{F} $ on $\ V $ such that $\ \mathcal{F}_{Y} =
\mathcal{T}_{Y} $ for all $\ Y \in \Sigma $.

On Minkowski space this topology coincides with the topology $\
Z$ defined by Zeeman mentioned above, for two specially chosen
systems $\ \Sigma $ which are significant for special relativity.
Since $\ \mathcal{T} $-open subset of $\ V $ always satisfies
condition ($\star $),  $ Z $ is always finer than $\ \mathcal{T} $.

Further G$\ddot{o}$bel defines a \emph{Special system} $\ \Sigma =
( \Gamma , \Delta )$ of $\ V $ as follows : \\ $\ \Sigma = (
\Gamma , \Delta )$ is called a special system of $\ V $ if there
is a locally finite covering $\ \mathcal{U} $ of $\ V $ by
neighbourhoods \emph{U}, such that $\ \Gamma =
\displaystyle\bigcup_{U \in \mathcal{U}} \Gamma_{U} $ and $\
\Delta = \displaystyle\bigcup_{U \in \mathcal{U}} \Delta_{U} $
where $\ \Gamma_{U} = \{ X \in \Gamma / X \subset U \} $ and $\
\Delta_{U} = \{ X \in \Delta  / X
\subseteq U \}$ have the following properties:\\
(i) If $\ X \in \Sigma = \Gamma \cup \Delta $ , then X is a closed
subset of $\ V $. \\
(ii) If $\ X \in \Gamma_{U} , Y \in \Gamma_{V} $ and $\ | X \cap Y
| = \infty $  then $\ X \cap V = Y \cap U $ for all $\ U, V \in
\mathcal{U}$. ( Here $\ |A|$ denotes cardinality of A) \\
(iii) If $\ p, q \in U \in \mathcal{U} $ and $\ \Gamma_{U} (p,q) =
\{ X \in \Gamma_{U} / p, q \in X \} $ is infinite, then $\ p = q
$.\\
(iv) We have $\ | X \cap Y | \leq 1$  for all $\ X \in \Gamma $
and $\ Y \in \Delta $. -------------  ( $\ \star \ \star $ )

With this definition, the following results follow:\\
\noindent\textbf{Proposition 4.1:} Let $\ \Sigma $ be a special system of $\
V $ and $\ f : [0, 1] \rightarrow V $ be a 1 - 1 map which is $\
\Gamma $- directed at $\ p \in f ([0,1])$. Then $\ f $ is a
piecewise $\ \Gamma $-curve at $\ p $ if $\ f $ is continuous at
$\ p $ with respect to the Zeeman topology Z.

(A curve $\ f $ is called $\ \Gamma $ - directed at $\ p \in f
([0,1])$ if there is a neighbourhood \emph{U} of $\ p $ defined by
( $\ \star \ \star $ ) such that if $\ p \neq q \in U \cap f
([0,1])$, then $\ \Gamma_{U}(p,q) \neq \phi $. $\ f $ is called a
piecewise $\ \Gamma $-curve at $\ p = f(a) $ if there are $\ b , c
$ with $\ 0 \leq b < a < c \leq 1 $ such that $\ f ([b,a])
\subseteq X $ and $\ f([a,c]) \subseteq Y $ for some $\ X,Y \in
\Gamma_{U} $) ).

\noindent\textbf{Proposition 4.2 :} If $\ \Sigma $ is a special
system of $\ V $ and $\ f $ is a Z-continuous curve which is $\
\Gamma $-directed at each point $\ p \in  f ([0,1])$, then $\ f $ is a piecewise $\ \Gamma $-curve.\\
This implies the following:

\noindent\textbf{Proposition 4.3 :} For a manifold $\ (V,
\mathcal{T})$ with an affine connection, following two statements
are equivalent:\\
(i) the curve $\ f $ is a piecewise geodesic i.e. $\ f $ is a
broken geodesic line with a finite number of edges.\\
(ii) the 1 - 1 map $\ f : [0,1] \rightarrow V $ is continuous with
respect to the Zeeman topology Z.

G$\ddot{o}$bel then restricts Zeeman topology on a space-time and
studies Zeeman topology by incorporating electromagnetic fields.
To state the results proved by G$\ddot{o}$bel in this situation,
we need to understand certain notations:

Let $\ V $ denote a space-time for general relativity and F be a
given electromagnetic field on $\ V $. An electric charge $\ q_{p}
$ of a test particle $\ p $ has its absolute value bounded by a
number depending on F, and mass $\ m_{p}$ of this particle ($\
m_{p} > 0$) is bounded by a number depending on the gravitational
field. Since the charge-spectrum Q and mass spectrum \emph{$\ W $}
are discrete, there are finitely many possible values $\ q_{p} \in
Q $ and $\ m_{p} \in W $ for test particles $\ p $. We assume the
presence of charge free test particles so that $\ O \in Q $. If Q
= {0}, we allow the mass spectrum $W$ to be arbitrarily $\ >  0 $.
Under these conditions, there are covering $\mathcal{U}$ and
$\mathcal{B}$ which are locally finite, so that there are only
finitely many world lines of freely falling test particles in $\ U
\in \mathcal{U} $ from $\ p \in U $ to $\ q \in U $ if $\ p \neq
q$. For $\ U \in \mathcal{U} $, let $\ \Gamma^{m}_{qU} $ be the
set of all world lines of freely falling test particles and let $\
\Delta_{U}$ be all closed space-like $\ C^{1} $-hypersurfaces of
$\ \overline{V}$. (Here $\ W \in \mathcal{B} $ such that there is
one and only one $\ U(W) \in \mathcal{U} $ which contains the
closure $\ \overline{W}$ of $\ W $). The corresponding system $\
\Sigma^{W}_{Q}$  is then a special system of $\ W $. Then the
following result holds :\\
\noindent\textbf{Proposition 4.4:} If $\ V $ is a space-time with
a given external electro-magnetic field F and a world line $\ f $,
the following statements are equivalent: \\
(a)  $\ f $ is continuous with respect to the Zeeman topology  $\
Z( \Sigma^{W}_{Q}, \mathcal{T} )$.\\
(b)  $\ f $ is a chain of finitely many connected world lines of
freely falling charged test particles.

If F = 0, then Z -continuous world lines are future directed
time-like piecewise geodesic lines. For simplicity, we denote $\ Z(
\Sigma^{W}_{Q}, \mathcal{T} )$ by $\ Z_{R}$. Then open sets with
respect to $\ Z_{R}$ are described as follows:\\
A subset Y of $\ V $ is open with respect to $\ Z_{R}$ iff $\ Y
\cap U $ is open in $\ (U, \mathcal{T}_{U})$ for the following
subsets $\ U $ of $\ V $ :\\
(I) \emph{U} is an arbitrary closed space-like hypersurface
contained in a simple region of $\ V $.\\
(II) \emph{U} is the world line of an arbitrary charged test
particle $\ p $ freely falling in the gravitational and the
electro-magnetic field within a simple region of $\ V $.\\
If Q = {0}, then condition (II) is equivalent to  \\
(II)' \ U is an arbitrary time-like geodesic in a simple region of
$\ V $. \\
If U is a simple neighbourhood of $p$ then let $\ U^{*}(p) = (U \setminus \mathcal{C}_{U}(p))\cup \{ p \}$.\\
\noindent\textbf{Lemma 4.5:} The set $\ U^{*}(p) $ defined above is a $\ Z(\sum^{W}_{Q}, \tau )$- neighbourhood of $p$.\\
G$\ddot{o}$bel then proves an important result that\\
\noindent\textbf{Proposition 4.6 :} The topology induced by $\
Z_{R}$ on a light cone is discrete. \\
Thus we do not have any geometric information along a light ray.\\
The main theorem of G$\ddot{o}$bel [7] is the following (which he
proves in the last section of his paper).

\noindent\textbf{Theorem 4.7:} Let h be a mapping from space-time
$\ V $ onto a space-time $\ V^{'} $. The following are equivalent:\\
(a)  h is a homeomorphism with respect to Zeeman topology $\ Z$. \\
(b)  h is a homothetic transformation.

Unusual property of Zeeman topology is that homeomorphism
characteristic of h implies its differentiability as well as its
`linearity', since h is an isometric map `up to scaling'. Thus we
can state this property in the following forms: \\
\noindent\textbf{Theorem 4.8: } The space-times $\ V $ and  $\
V^{'}$ are homeomorphic with respect to Zeeman topology if and
only if they are isometric (up to a constant positive factor).\\
\noindent\textbf{Theorem 4.9:}  The group of all homeomorphisms
with respect to the Zeeman topology coincides with the group of
all homothetic transformations of space-time $\ V $ onto itself.\\
Thus Zeeman topology contains all information about the metric.

We again note here that (locally) causal maps defined by G$\ddot{o}$bel [7] in Section 2 and described in Section 5 are
similar to causal maps of Garc$\acute{i}$a-Parrado and Senovilla [23], and subsequently similar to K-causal maps described and studied by Sujatha Janardhan and R.V.Saraykar [31].

As far as Minkowski space-time is concerned, Zeeman [1] has suggested other topologies on it. G$\ddot{o}$bel generalized some of the results which hold for these topologies. Following remarks are in order about these topologies :\\
\textbf{Remark 1} : The topology $Z_1$ defined by Zeeman is now well-known as t-topology studied by Nanda [4]. The induced topology on any space axis is discrete. Under this topology, G$\ddot{o}$bel has generalized this result as follows :\\
\noindent\textbf{Theorem 4.10:} Let $\ f : I \rightarrow (V,
Z_{1})$ be a continuous map of the unit interval I into $\ V $
(endowed with $\ Z_{1}$- topology). If $\ f $ is strictly order
preserving, i.e. $\ x < y $ implies $\ f(x) < f(y)$ (i.e. the
vector $\ f(y) - f(x) $ is time like), then the image $\ f(I) $ is
a piecewise linear path, consisting of a number of intervals along
time axis. \\
Further, this topology has a physically attractive feature as
follows:\\
If $\ f : I \rightarrow V $ be an embedding (not necessarily order
preserving), then $\ f(I) $ is a piecewise linear path along time
axes, zig-zagging with respect to time orientation like the
Feynman track of an electron.\\
Hawking, King and Mc Carthy [9] has defined \emph{Feynman path} mathematically precisely as follows: \\
Let $\ K(p, \texttt{U})$ denote $\ I^{+}(p, \texttt{U}) \cup
I^{-}(p, \texttt{U}) \cup \{p \} $ where \texttt{U} denotes an
open convex normal neighbourhood of $\ p $. A path $\ \gamma : I
\rightarrow  V $ is a \emph{Feynman path} if $\ \gamma $ is
continuous and for each $\ t_{0} \in I$, there is an open
connected neighbourhood \emph{U} of $\ t_{0} $, and an open convex
normal neighbourhood \texttt{U} of $\ p = \gamma(t_{0})$ such that
$\ \gamma(U) \subseteq K (p, \texttt{U})$.\\
A locally one - one Feynman path is then a Feynman track mentioned
above.\\
Let G denote the group of automorphisms of $\ V $ given by \\
(i) the Lorentz group of all linear maps leaving quadratic form Q
invariant \\
(ii) translations and \\
(iii) dilatations.\\
Every element of G either preserves or reverses the partial ordering `$\ < $' mentioned above. These features have been studied in details by Nanda, Dossena and Kim.\\
\textbf{Remark 2} : The topology $Z_2$ defined by Zeeman is well-known as s-topology studied by Nanda [3,4]. The induced topology on any time axis is discrete. Homeomorphism group of this topology was determined by Nanda thus proving another version of Zeeman conjecture. Topological properties of t-topology and s-topology have been studied by  G. Agrawal and S. Shrivastava [14,15 ] as mentioned in Section 2.\\
\textbf{Remark 3} : The topology $Z_3$ defined by Zeeman has been studied by Williams[6]. This topology is denoted by $M^F$ in Section 2. Thus we note that $\ Z_{3} $ is then coarser than $\ Z $ but finer than the Euclidean topology. This topology has the following properties
which are  possessed by $\ Z $: \\
(a)  It is not locally homogeneous and the light cone through any point can be deduced from it.\\
(b)  The group of all homeomorphisms with respect to $\ Z_{3} $ is generated by inhomogeneous Lorentz group and dilatations.\\
(c)  It induces the 3- dimensional Euclidean topology on every space axis and the 1 - dimensional Euclidean topology on every time axis.

However, this topology does not satisfy the theorem mentioned above. Nevertheless, the group of homeomorphisms of $\ (V, Z_{3})$ is G. Thus although $\ (V, Z_{3})$ has a countable base of neighbourhoods for each point, it is physically less attractive than $\ Z $. Such topologies can also be described on a general space-time following G$\ddot{o}$bel's method.

Ulf Lindstrom [11] re-examined the separating topology studied in earlier works. Using methods and ideas in papers by G$\ddot{o}$bel, Hawking, King and McCarthy, he introduced  a new class of topologies $\ \{ S_{nm} \}$. The topology $\ \{ S_{nm} \}$ is the finest which induces Euclidean topology on time-like $\ C^{n} $- and space-like $\ C^{m} $ -curves.  A relation between $\ \{ S_{nm} \}$  and some topologies studied by G$\ddot{o}$bel is derived- For an arbitrary space-time the group of homeomorphisms is shown to be the smooth conformal diffeomorphism group. The restriction to strongly causal space-times employed in earlier work is no longer necessary.
We note that Lindstrom topology reduces to Williams $\ C^1 $ topology $\ M^{F}$ for $\ m = n = 1$ on Minkowski space. Group of $\ C^1$ homeomorphisms is $\ C^1$ conformal diffeomorphisms as noted in Section 2.

Finally, we add a comment about the work of Mashford [19] : As is well-known, a space-time in the general theory of relativity is a Lorentz manifold modeled on 4-dimensional Euclidean space, which is locally a Minkowski space. Mashford [19] constructs a tangent bundle whose base space is not a Lorentz manifold, but is a set $Y$ of events which is equipped with an acyclic signal relation $\ \sim \rightarrow $ and the $\ \sim \rightarrow $  structure of $Y$ is locally that of Minkowski space with Zeeman topology. Moreover, the \emph{piecing together} maps are smooth in an appropriate sense. The parent space $E$ is the tangent bundle $TY$ of $Y$. Mashford then proves that this bundle has, as structure group, the group of linear causal automorphisms of Minkowski space, which coincides with the group $\ G$ of Lorentz transformations along with translations and dilatations which has been discussed in Section 2.

\section{Conclusion}

In this article, we have given a short review of Zeeman- and Zeeman -like fine topologies on Minkowski space and space-time of general relativity. We have avoided giving detailed proofs of the results mentioned, otherwise the article would have become lengthy. To the best of our knowledge, we have reviewed most of the research work which appeared on this topic since the first paper was published by Zeeman in 1967. To get a consolidated view about definitions and the main properties of these topologies like their homeomorphism groups and topological properties, we give  two tables summarizing definitions and their properties :\\
Definitions and Properties of Fine topologies on Minkowski space (refer Table 1) and \\
Fine topologies on  space-times of general relativity (refer Table 2)
\begin{table}[h]
\caption{\textbf{Definitions and Properties of Fine topologies on Minkowski space}}
\begin{tabular}{|p{.5cm}|p{5cm}|p{3cm}|p{5cm}|}
  \hline
  Sr. No & Fine topology & Homeomorphism Group & Topological properties  \\  \hline
  1 & Zeeman topology $\ M^{Z} $ (1967): Finest topology which induces  three dimensional Euclidean  topology on every space-axis and one dimensional Euclidean topology on every time-axis & G = Lorentz group with translations and dilatations  & Dossena (2007) : Neither locally compact nor Lindelof, not normal, separable but not first countable, path-connected but not simply connected  \\ \hline
  2 & s-topology $\ M^{s} $ :Nanda (1971) : Finest topology which induces three dimensional Euclidean topology on every space-like hypersurface & G & G.Agrawal and S. Shrivastava (2012) : separable, first countable, path-connected, not regular, not metrizable, not second countable, noncompact, and non-Lindelof, not simply connected \\ \hline
  3 & t-topology $\ M^{t} $ : Nanda (1972) : Finest topology which induces one dimensional Euclidean topology on every time-like line & G & G.Agrawal and S. Shrivastava (2009) : separable, first countable, path-connected, not regular, not metrizable, not second countable, not locally compact, not simply connected \\ \hline
  4 & A-topology $\ M^{A} $ : Nanda (1979) : Finest topology which induces one dimensional Euclidean topology on every time-like line and light-like line and three dimensional Euclidean topology on every space-like hypersurface & G & G.Agrawal and Soami Pyari Sinha (2014) : separable, not first countable, connected and path-connected, not normal, not metrizable, Not comparable with t-topology nor with s-topology \\ \hline
  5 & Fine topologies $\ M^{F} $  by Williams (1974): Finest topology which induces one dimensional Euclidean topology on every time-like line and space-like line & Conformal group of Minkowski space whose $\ C^{1} $ subgroup is G & Hausdorff, separable, first countable, but not regular and hence not metrizable \\ \hline
  6 & $\ M^{L} $ : Finest topology which induces one dimensional Euclidean topology on every straight line & $\ C^{1} $ homeomorphisms form projective group generated by full linear group and translations & Weaker than $\ M^{F} $ and $\ M^{T} $ , Hausdorff, separable and first countable, not regular and hence not metrizable  \\ \hline
\end{tabular}
\end{table}
\begin{table}[h]
\caption{\textbf{Fine topologies on  space-times of general relativity}}
\begin{tabular}{|p{.5cm}|p{5cm}|p{3cm}|p{5cm}|}
  \hline
  Sr. No & Fine topologyon space-time of GR & Diffeomorphism Group & Topological properties  \\  \hline
  1 & HKM-path topology described by Hawking-King-McCarty [1976] & Conformal diffeomorphisms  &  Hausdorff, path connected and locally path connected, first countable, separable, but not normal or locally compact  \\ \hline
  2 & Extended HKM-topology (Kim - 2006 )& Conformal isomorphism group & Finer than Alexandrov topology \\  \hline
  3 & S-topology on Lorentz manifolds (Domiaty 1985 )& Conformal $\ C^{\infty}$-diffeomorphisms & Hausdorff, first countable and separable, not regular and hence not metrizable, path connected and locally path connected \\  \hline
  4 & Zeeman -like fine topology in general relativity described by G$\ddot{o}$bel (1976) & Homeomorphism group with respect to Zeeman-like topology is the group of all homothetic transformations of V & Strongly causal space-times \\  \hline
  5 & Lindstrom (1978) : Finest topology $\ S_{nm} $ that induces the topology as a submanifold on time-like $\ C^{n} $-curves and on space-like $\ C^{m} $-curves & Group of Conformal $\ C^{\infty} $ -diffeomorphisms or group of all homothetic transformations of V & Space-time need not be strongly causal \\  \hline
\end{tabular}
\end{table}
Whereas fine topologies have interesting topological properties and their homeomorphism groups are physically useful, it is nevertheless true that manifold structure is not compatible with fine topologies. This is because, topologically, a manifold is second countable, Hausdorff and paracompact, and hence normal and metrizable, whereas fine topologies are not, in general, normal (and hence not metrizable).  Moreover, it is also true that unless differential structure is there, we can not define notions of connection and curvature and hence fine topologies may not be useful in discussing Einstein field equations in general theory of relativity.
Finally, we would like to refer to a paper by A. Heathcote [35] , where it has been argued that the suggestions for replacement of manifold topology with fine topology misrepresent the significance of the manifold topology and overstate the necessity for a finer topology. He claims to have given a realist view of space-time topology. Other philosophical issues about space-time have been discussed by D. Dieks and M. Redel in two volumes [36,37].\\

\section{References}
1. Zeeman E.C. : Topology ,6, 161 (1967)\\
2. Dossena G. : Jour. Math. Phys., 48, 113507 (2007)\\
3. Nanda S. : Jour. Math. Phys., 12, 394 (1971)\\
4. Nanda S. : Jour. Math. Phys., 13, 12 (1972)\\
5. Nanda S. : Jour. Austral. Math. Soc., 21(Series B), 53 (1979)\\
6. Williams G. : Proc. Camb. Phil. Soc. 76, 503 (1974)\\
7. G$\ddot{o}$bel R.: Commun. Math. Phys. 46, 289 (1976).\\
8. G$\ddot{o}$bel R., Jour. Math. Phys. 17, 845 (1976).\\
9. Hawking, S.W., King. A.R., McCarthy P.J. : Jour. Math. Phys. , 174 (1976).\\
10. Malament, D.B.: Jour. Math. Phys. 18, 1399 (1977).\\
11. Lindstrom U.: Further considerations on the separating topology for the space-times of general relativity,
    http://www.iaea.org/inis/collection/NCLCollectionStore/Public/09/374/9374573.pdf. \\
12. Popvassilev S.G.: Mathematica Pannonica, 5/1, 105 (1993). \\
13. Kim D.H.: Jour. Math. Phys. 47, 072503 (2006).\\
14. Agrawal G. and Shrivastava S. : Jour. Math. Phys. 50, 053515 (2009). \\
15. Agrawal G. and Shrivastava S.: ISRN Math. Phys. Article ID 896156 (2012). \\
16. Agrawal G. and Soami Pyari Sinha, Properties of A- topology, Preprint (2014). \\
17. Low R. : Class. and Quant. Grav. 27(10), 107001 (2010 ).\\
18. Fullwood D.T.: Jour. Math. Phys. 33, 2232 (1992).\\
19. Mashford J.S. : Jour. Math. Phys. 22, 1990 (1981).\\
20. Penrose R. : Techniques of differential topology in relativity, SIAM, Philadelphia,1972, (Regional Conf. Series in Appl. Math., vol. 7).\\
21. Ehlers, J.: General relativity and kinetic theory. In: Sachs,R.K. (Ed.): Relativita generale a cosmologia. New York: Academic Press 1971.\\
22. Joshi P.S.: Global aspects in gravitation and cosmology, Clarendon Press, Oxford, 1993.\\
23. Garcia-Parrado A. and Senovilla J M : Class. Quant. Grav. 20, 625 (2003). \\
24. Sujatha Janardhan and Saraykar R.V.: Gravitation and Cosmology, 19, 42 (2013). \\
25. Nanda S. and Panda H.K. : Inter. Jour. Theor. Phys. 12(6), 393 (1975) \\
26. Struchiner I. and Rosa M.A.F., arXiv:math-ph/0504071v1 (2005).\\
27. Domiaty, R.Z. : Gen. Relat.Grav. 17(12), 1165 (1985).\\
28. Domiaty, R.Z. : Gen. Top. Appl. 20, 39 (1985).\\
29. Budic R. and Sachs R.K.: J. Math. Phys. 15, 1302 (1974).\\
30. Huang W.: Jour. Math. Phys., 39, 1637 (1998).\\
31. Peleska J.: Aequationes Mathematicae, 27, 1637 (1998).\\
32. Sorkin R.D. and Woolgar E.: Class. Quant. Grav. 3, 1971 (1996).\\
33. Sujatha Janardhan and Saraykar R.V.: Pramana, a jour. of Physics,  70(4), 587 (2008). \\
34. Minguzzi E.: Commun.Math.Phys. 290, 239 (2009).\\
35. Heathcote A. , British Jour. for Phil. Of Sci., Vol.39, Issue 2, 247-261 (1988)\\
36. Dieks D. and Redel M. ,  Vol. 1, Philosophy and Foundations of Physics Series,  Elsevier, (2006).\\
37. Dieks D. and Redel M. ,  Vol. 2, Philosophy and Foundations of Physics Series,  Elsevier, (2008).
\end{document}